# Value of Information Analysis for External Validation of Risk Prediction Models

Mohsen Sadatsafavi[1], Tae Yoon Lee[1], Laure Wynants[2,3], Andrew Vickers[4], Paul Gustafson[5]

1. Respiratory Evaluation Sciences Program, Faculty of Pharmaceutical Sciences, University of British Columbia, Vancouver, British Columbia, Canada
2. Department of Epidemiology, CAPHRI Care and Public Health Research Institute, Maastricht University, Maastricht, The Netherlands
3. Department of Development and Regeneration, KU Leuven, Leuven, Belgium
4. Department of Epidemiology and Biostatistics, Memorial Sloan Kettering Cancer Center, New York, New York, USA
5. Department of Statistics, University of British Columbia, Vancouver, British Columbia, Canada

**Corresponding author:**    Mohsen Sadatsafavi
Room 4110, 2405 Wesbrook Mall,
Vancouver, BC, V6T1Z3, Canada
Tel: +1 604 827 3020
Email: msafavi@mail.ubc.ca

Table count: 1
Figure count: 5
Word count: 4,484 (without abstract, references, figures, and tables)

**Short title:** Value of Information and Validation of Risk Prediction Models

**Source of support:** This study was supported by the Canadian Institutes of Health Research (PHT 178432).

**Conflict of interest:** None declared

**ABSTRACT**

**Background**: Before being used to inform patient care, a risk prediction model needs to be validated in a representative sample from the target population. The finite size of the validation sample entails that there is uncertainty with respect to estimates of model performance. We apply value-of-information methodology as a framework to quantify the consequence of such uncertainty in terms of NB.

**Methods**: We define the Expected Value of Perfect Information (EVPI) for model validation as the expected loss in NB due to not confidently knowing which of the alternative decisions confers the highest NB at a given risk threshold. We propose methods for EVPI calculations based on Bayesian or ordinary bootstrapping of NBs, as well as an asymptotic approach supported by the central limit theorem. We conducted brief simulation studies to compare the performance of these methods, and used subsets of data from an international clinical trial for predicting mortality after myocardial infarction as a case study.

**Results**: The three computation methods generated similar EVPI values in simulation studies. In the case study, at the pre-specified threshold of 0.02, the best decision with current information would be to use the model, with an expected incremental NB of 0.0020 over treating all. At this threshold, EVPI was 0.0005 (a relative EVPI of 25%). When scaled to the annual number of heart attacks in the US, this corresponds to a loss of 400 true positives, or extra 19,600 false positives per year, indicating the value of further model validation. As expected, the validation EVPI generally declined with larger samples.

**Conclusion**: Value-of-information methods can be applied to the NB calculated during external validation of clinical prediction models to provide a decision-theoretic perspective to the consequences of uncertainty.

**HIGHLIGHTS**

- External validation is a critical step before recommending a risk prediction model for patient care, but the finite size of the validation sample creates uncertainty about the performance of the model in the target population.
- In decision theory, such uncertainty is associated with loss of net benefit because it can prevent one from identifying whether the use of the model is beneficial over alternative strategies.
- We define the Expected Value of Perfect Information for model validation as the expected loss in net benefit by not confidently knowing if the use of the model is net beneficial.
- Value-of-Information methods can help determine the value of further validation studies.

**INTRODUCTION**

Once a risk prediction model is developed, its performance needs to be examined in an independent sample obtained from the target population of interest. Such external validation is considered a prerequisite for the model to move to the next stage, whether a clinical impact study or clinical implementation(1). During external validation, model performance is typically examined in terms of calibration, discrimination, and net benefit (NB)(2). The NB is a decision-theoretic metric that considers the benefits and harms associated with risk stratification(3). Due to its deep roots in decision theory as well as its ease of calculation, the NB approach has become a widely used tool for the evaluation of prediction models.

Given the finite size of the external validation sample, the assessment of the performance of a risk prediction model is accompanied by uncertainty. This is typically approached as a statistical inference problem (e.g., by presenting error bands around the calibration plot, or 95% Confidence Interval [CI] around the c-statistic). Recent works on power and sample size calculations for external validation studies have proposed targeting pre-specified standard errors on overall calibration, calibration slope, c-statistic, and NB(4). However, there is a current debate on the relevance of inference around NB, a measure of clinical utility, using standard inferential methods such as confidence intervals(5,6).

As risk prediction models are ultimately used to inform patient management, uncertainty in their performance can be assessed in terms of its impact on the outcomes of medical decisions. From this perspective, the finite size of the validation sample can lead to incorrect conclusions, for example, recommending the use of the model where in fact the best strategy is to treat all eligible patients. Thus, conclusions based on a finite validation sample can be associated with loss of clinical utility. However, the use of standard statistical methods such as confidence intervals to express the uncertainty of a decision-analytic measure is highly questionable(5,6).

Value-of-information (VoI) analysis is a set of concepts and methods rooted in decision theory that aims at quantifying the expected loss due to uncertainty in decisions(7). In a recent work, we applied such methods to the development phase of risk prediction models(8). We defined the Expected Value of Perfect Information (EVPI) for model development as the expected loss in NB due to uncertainty in the coefficients of a prediction model developed based on a finite sample. In this paper, we extend such a concept from the development to the validation phase, and propose the validation EVPI as the expected loss of clinical utility due to uncertainty about the performance of the model inferred from a finite validation sample.

**Context**

We focus on a previously-developed risk prediction model for a binary outcome that is now undergoing external validation in a representative sample of $n$ subjects form a new target population. In this context, the model can be seen as a function $\pi(X)$ that maps the covariate pattern $X$ to predicted risk $\pi$ for a binary outcome $Y$ (e.g., risk of in-hospital mortality due to sepsis, or experiencing an asthma attack in the next 12 months). If there are other competing models during external validation, they can also be considered in this framework. However, without loss of generality and for the sake of brevity, in what follows we assume only one model is being assessed in external validation. As is implicit in NB calculations, we assume the decision-maker is risk neutral and the only source of evidence is the validation sample at hand. All the analyses are conducted in the R statistical programming environment(9).

The measure of clinical utility that we will focus on is NB(3). In brief, to turn a continuous predicted risk to a binary classification (e.g., low- versus high-risk) to inform a treatment decision, a decision-maker needs to specify a context-specific threshold $z$ by considering the benefits and harms of the treatment(10). Treatment is offered to patients whose predicted risk is above this threshold. If the predicted risk is precisely at this threshold, the decision-maker would be ambivalent between treatment and no treatment. Such ambivalence implies that the decision-makers assigns a relative weight of $z/(1-z)$ to each false positive classification compared to each true positive classification(3). Consequently, the NB of the model at threshold $z$, compared with treating no one, can be calculated in net true positive units as:

$$NB_{model} = P_{True\ Positive}(z) - P_{False\ Positive}(z)\frac{z}{1-z}.$$

For brevity, we drop the notation that indicates the left side is dependent on $z$. Here, $P_{True\ Positive}(z) = P(\pi \geq z, Y = 1)$ and $P_{False\ Positive}(z) = P(\pi \geq z, Y = 0)$. A consistent estimator of this NB in the sample is

$$\widehat{NB}_{model} = \frac{1}{n}\sum_{i=1}^{n} I(\pi_i \geq z)\{Y_i - (1 - Y_i)\frac{z}{1-z}\}.$$

The strategy of using the model for patient management competes with at least two 'default' strategies: treating no one, and treating all. The NB of the former is zero by definition. The NB of treating all is

$$NB_{all} = P_0 - (1 - P_0)\frac{z}{1-z},$$

where $P_0$ is the outcome prevalence in the population. $NB_{all}$ can be consistently estimated in the sample as

$$\widehat{NB}_{all} = \frac{1}{n}\sum_{i=1}^{n}\{Y_i - (1 - Y_i)\frac{z}{1-z}\}.$$

**Motivating example: prediction of mortality after acute myocardial infarction (AMI)**

Identifying the risk of immediate mortality after an AMI can enable stratification of more aggressive treatments for high-risk individuals. GUSTO-I was a large clinical trial of multiple thrombolytic strategies for AMI(11). This dataset has frequently been used to study methodological aspects of developing or validating risk prediction models(12–14). In line with a previous study, we used the non-US sample of GUSTO-I (n=17,796) to fit a prediction model for 30-day post-AMI mortality, and the US sample (n=23,034) for validating it(15). Such a validation sample is larger than typical sizes of samples in most practical contexts. To make a case for our developments, we assume that we have access to data for only 500 patients; we will later use the entire sample to study how our proposed metric changes with sample size. We randomly selected, without replacement, 500 individuals from the entire US sample to create such an exemplary validation dataset. 30-day mortality was 7.0% in the non-US sample, 6.8% in the entire

US sample, and 8.6% in the validation sample. As in previous case studies using these data(8), our primary threshold of risk interest is 0.02, above which more aggressive thrombolytic treatments are justified.

Our candidate risk prediction model is similar to the previously developed ones using this dataset(8). We did not apply any shrinkage given the large development sample. The final model for 30-day post-AMI mortality based on applying logistic regression to the entire development sample was

$$logit(\pi) = -2.084 + 0.078 * age + 0.403 * [AMI\ location\ other] + 0.577 * [Anterior\ AMI] + 0.468 * [Previous\ AMI\ history] + 0.767 * [Killip\ score] - 0.077 * \min([blood\ pressure], 100) + 0.018 * pulse.$$

The c-statistic of this model is 0.847 in this validation sample. **Figure 1** is the "decision curve" that depicts the empirical NB of the model ($\widehat{NB}_{model}$, blue) alongside those of treating none and treating all. At the 0.02 threshold, the NB of the model was 0.0692, whereas the NB of treating all was 0.0672. As such, based on this validation sample, the use of the model confers an incremental NB of 0.0020 over the next best decision, which is to treat all.

**Figure 1:** Decision curve ($\widehat{NB}$) of the candidate model (blue), treating all (black oblique line), and treating none (black horizontal line) in the validation sample A bootstrapped 95% confidence interval is given for the candidate model and treating all strategies (gray curves).

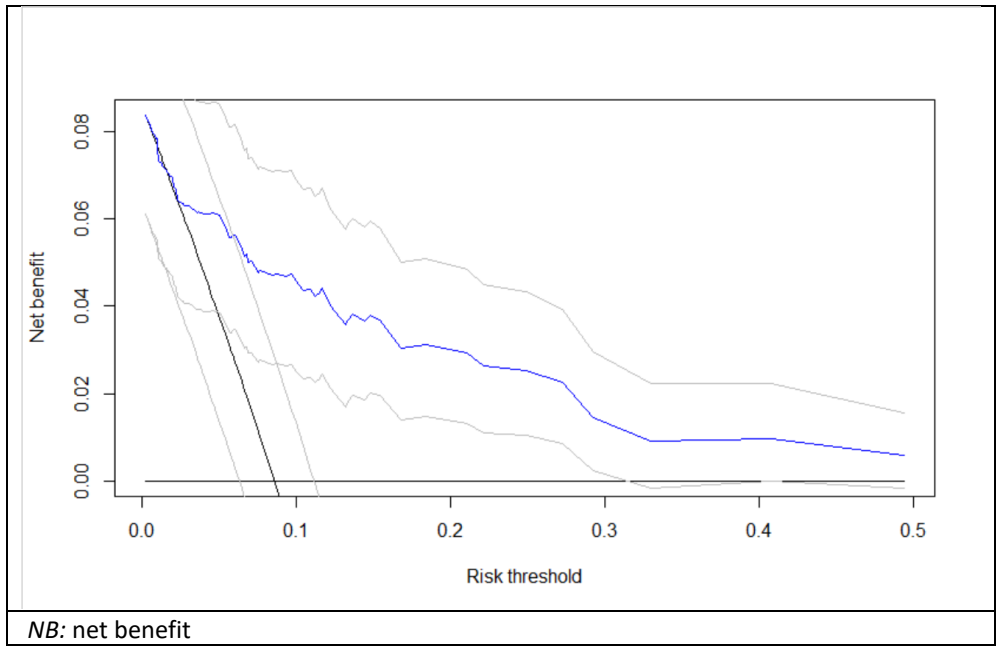

*NB:* net benefit

The gain in NB by using the model over the default strategies can be presented either in terms of change in the number of true positives while holding the number of false positives constant, or vice versa(3). Here, the difference of 0.0020 means for every 1,000 treatment decisions, the use of the model will result in an expected gain of 2 true positives. Because a 0.02 threshold implies an exchange rate of 49 between false and true positives, this can also be interpreted as the use of the model resulting in, on average, 2 more true positives × 49 = 98 fewer false positives (i.e., fewer unnecessary treatments) per 1,000 decisions.

However, due to the finite validation sample, there is uncertainty in which strategy confers the highest benefit in the population. Vickers et. al. proposed bootstrapping for inference on NBs(16). In this scheme, a bootstrap sample from the validation data is obtained, and $\widehat{NB}_{model}$ and $\widehat{NB}_{all}$ are calculated at thresholds of interest. Repeating this step many times provides an empirical distribution for NBs that can be used for constructing confidence intervals around NBs or their differences. Using the percentile method applied to 10,000 bootstraps, the 95%CIs for $\widehat{NB}_{model}$ and $\widehat{NB}_{all}$ are presented as gray curves in **Figure 1.** We also obtain a 95%CI for the $\widehat{NB}_{model} - \widehat{NB}_{all}$ of -0.0050 to +0.0055 at the 0.02 threshold, which indicates the evidence is also compatible with treating all being the best strategy.

**A Bayesian approach towards interpreting uncertainties around NB**

The conventional bootstrap is akin to assigning a random weight to each observation, with weights drawn from a scaled multinomial distribution. Rubin proposed the Bayesian bootstrap, where weights are instead generated from a Dirichlet distribution(17). They showed that a summary statistic derived from such a bootstrapped sample can be interpreted as a random draw from the posterior distribution of the corresponding population parameter given the sample and a non-informative prior on the underlying data-generating mechanism(17). The similarity of weighting scheme and numerical results between the ordinary and Bayesian bootstraps has resulted in the former being also interpreted in a Bayesian view, as in the approximate Bayesian bootstrap method for the imputation of missing data(18), or in VoI analysis of cost-effectiveness trials(19).

Such Bayesian interpretation of the bootstrap enables us to make probabilistic statements about the true NBs. For example, we can count the proportion of bootstraps in which the NB of the model is higher than the NB of the default strategies. This quantity, proposed by Wynants et al and termed *P(useful)*, will be the posterior probability that the model-based treatment is truly the best strategy(20). ***Figure 2*** depicts the bootstrap distribution of $\widehat{NB}_{model} - \max\{0, \widehat{NB}_{all}\}$ based on 10,000 bootstraps; *P(useful)*, corresponding to the area that lies on the right side of 0, is 75.9%.

***Figure 2:*** Histogram of the incremental NB of the model ($\widehat{NB}_{model} - \max\{0, \widehat{NB}_{all}\}$) based on 10,000 bootstraps.

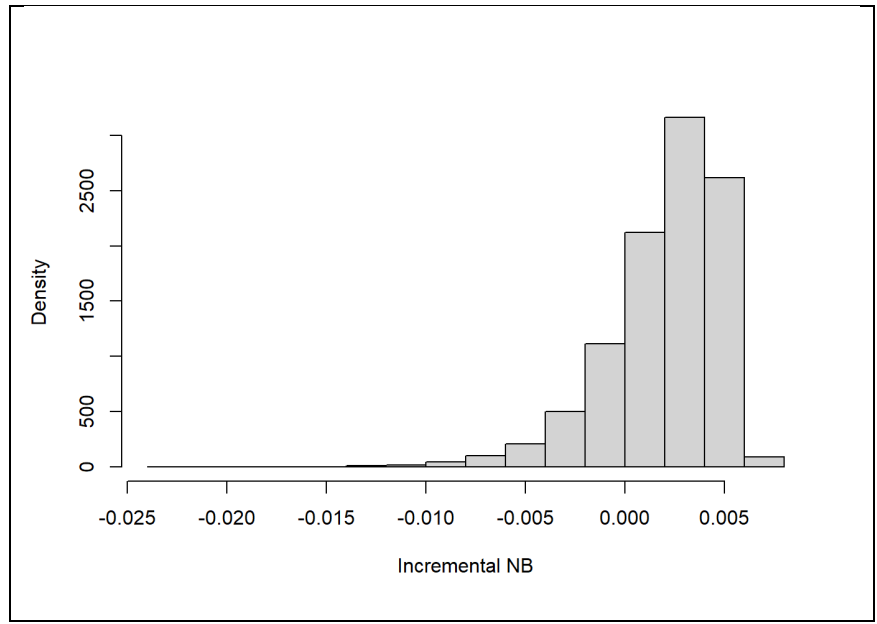

**The expected value of perfect information (EVPI)**

A *P(useful)*<1 indicates the possibility that our model, despite having the highest expected NB in this validation sample, might not truly be the best one, in which case the use of the model will be associated with a loss of NB. To quantify the magnitude of this loss, we can follow the steps demonstrated in **Table 1**. Each row in this table corresponds to a bootstrapped estimate of $NB_{model}$ and $NB_{all}$, which we can interpret as random draws from the posterior distribution of their true counterparts. In each row, we ask the question "*What decision would have we made, had we known these are the true NBs?*" In the first row, $NB_{model} > NB_{all} > 0$. As such, using the model remains the best strategy. However, in the second row, $NB_{all} > NB_{model} > 0$, so we would have chosen the strategy of treating all. Given the NBs in this row, this strategy would give us an extra NB of 0.0074 over the use of the model (last column).

| Table 1: Stepwise calculations for the expected gain with perfect information for the case study ||||||
| Iteration | $NB_{model}$ | $NB_{all}$ | Winning strategy | Gain in NB over current strategy (use the model)* |
|---|---|---|---|---|
| #1 | 0.0830 | 0.0801 | Use the model | 0 |
| #2 | 0.0936 | 0.1010 | Treat all | 0.0074 |
| #3 | 0.0787 | 0.0784 | Use the model | 0 |
| … | | | | |

| #10,000 | 0.0704 | 0.0654 | Use the model | 0 |
| **Average** | **0.0694** | **0.0674** | Use the model: **75.9%**[#] <br> Treat all: **24.1%** <br> Treat none: **0%** | **0.0005** |

Each row pertains to a random draw from the posterior distribution of $P(NB_{model}, NB_{all})$, the posterior distribution of true net benefits. For each row, we find the optimal strategy as the one with the highest benefit. The last column records the incremental net benefit of the winning strategy compared with the use of the model.

*Current strategy is to use the model because it has the highest expected NB (the last row)

#Corresponding to a P(useful) of 75.9%

*NB:* net benefit

Across the bootstraps, the average of $\widehat{NB}_{model}$ is 0.0694, higher than both 0 and the average of $NB_{all}$ (indeed, the averages of $\widehat{NB}_{model}$ and $\widehat{NB}_{all}$ across the bootstraps converge to their corresponding estimate in the original sample as the number of bootstraps grow); thus, in our Bayesian approach using the model is the best strategy based on the current evidence. However, in 24.1% of rows we would have declared treating all as the best strategy, had we known the true NBs, and in doing so would have gained clinical utility over the use of the model. We do not know which value of NBs will emerge as the truth, but we can probabilistically combine all scenarios by averaging the gain in NB across rows, which evaluates to 0.0005 (the average of last column values in **Table 1**). This is the expected NB loss due to the finite validation sample, or the expected NB gain by knowing the true NBs, a term that we call the Expected Value of Perfect Information (EVPI) for model validation.

Formalizing the above derivations, we quantify the EVPI by contrasting the expected NBs of the decision-making process under two scenarios of current information (estimating NBs with uncertainty in the sample), and perfect information (knowing the true NBs). With the validation sample at hand, the best we can do is to choose the strategy with the highest expected NB, an approach that confers an expected NB of $\max\{0, \mathrm{E}NB_{model}, \mathrm{E}NB_{all}\}$, with the expectation being with respect to $P(NB_{model}, NB_{all})$. On the other hand, if we know the true values of NBs, we could choose the most beneficial strategy with certainty, an approach that would confer a NB of $\max\{0, NB_{model}, NB_{all}\}$. We do not know the true value of NBs, but we can quantify the expected value of such an optimal strategy given our current knowledge, which would be $\mathrm{E}\{\max\{0, NB_{model}, NB_{all}\}\}$, with the expectation being, again, with respect to $P(NB_{model}, NB_{all})$. The validation EVPI is the difference between these two terms:

$$EVPI = \mathrm{E}\{\max\{0, NB_{model}, NB_{all}\}\} - \max\{0, \mathrm{E}NB_{model}, \mathrm{E}NB_{all}\}.$$

**How can we interpret the EVPI?**

EVPI is a non-negative scalar quantity in the same units as the NB of risk prediction models, with higher EVPI values indicating higher expected NB loss due to uncertainty. Given that the EVPI is in NB units, the consequence of uncertainty can be presented in the same way as the results of decision curve analysis. An EVPI of 0.0005 indicates that removing uncertainty about which strategy is the most beneficial is associated with an expected gain of 0.5 in true positives, or avoiding an expected 24.5 false positives (unnecessary treatments), for every 1,000 treatment decisions.

Theoretically, any EVPI>0 indicates potential value of future validation studies. However, a very low positive EVPI value would indicate a low yield from such a study. A given value of EVPI cannot be declared low or high without considering the decision context. EVPI measures the expected NB loss per treatment decision due to uncertainty, a loss that is potentially preventable by performing more validation studies. The true magnitude of this preventable loss is affected by the number of times the decision of interest is being made in the target population. For example, more than 800,000 AMIs occur every year in the US(21), and a guideline panel in charge of making a national-level recommendation for AMI treatment can consider our candidate model potentially applicable to all such events. In this case, the impact of validation uncertainty is equal to missing proper intervention in 400 true positive cases (patients with AMI who will die within 30 days), or imposing unnecessary treatments to 19,600 false positive cases (patients who will survive) per year. As validation studies often involve secondary use of data and thus are not overly expensive, procuring more samples to reduce this preventable loss seems justifiable.

An alternative way to contextualize an EVPI value, applicable when the model has the highest expected NB with current information, is the previously proposed relative EVPI(8). The relative EVPI compares the expected NB gain with perfect information with the expected NB gain due to use of the model. In our case study, the incremental NB of the model over the next best decision (treating all) is 0.0020. With perfect information, we expect to gain an extra NB of 0.0005. Thus, we can gain on average, 0.0025/0.0020=1.25, or 25% more efficiency by removing validation uncertainty. Such relative EVPI can thus be defined as(8):

$$rEVPI = \frac{\mathrm{E}\{\max\{0, NB_{model}, NB_{all}\}\} - \max\{0, \mathrm{E}NB_{all}\}}{\max\{0, \mathrm{E}NB_{model}, \mathrm{E}NB_{all}\} - \max\{0, \mathrm{E}NB_{all}\}}.$$

**Computation algorithms**

*(Bayesian) bootstrapping*

As explained earlier, a Bayesian interpretation of the bootstrap enables us to use this method readily for EVPI calculations. In this scheme, we interpret $(\widehat{NB}_{model}, \widehat{NB}_{all})$ from a Bayesian or ordinary bootstrapped sample as a random draw from $P(NB_{model}, NB_{all})$. This in turn enables the calculation of both terms for EVPIs via straightforward Monte Carlo simulations. This results in the following algorithm for EVPI (**Box 1**).

---

**Box 1:** Bootstrap-based EVPI computation*

1. Calculate the predicted risk for each individual in the validation sample.

2. For I = 1 to N (a sufficiently large number)

   2.1 Create data set $D^*$ as the (Bayesian) bootstrap of the validation data set. $D^*$ is a random from the posterior distribution of the population given the sample. This can be done via assigning a vector of weights W to each observation, with weights coming from $(W \sim Dirichlet(1,1,...,1)\#)$, or for ordinary bootstrap from $W \sim Multinom(n; 1/n, 1/n, ..., 1/n)$.

   2.2 Estimate $\widehat{NB}^*_{model}$ and $\widehat{NB}^*_{all}$ at thresholds of interest from $D^*$ and store their values.

   2.3. Let $maxNB^* = \max\{0, \widehat{NB}^*_{model}, \widehat{NB}^*_{all}\}$

---

3. Calculate $EVPI = \text{mean}(maxNB^*) - \max\{0, \text{mean}(\widehat{NB}^*_{model}), \text{mean}(\widehat{NB}^*_{all})\}$.

In calculating the expected NB under current information, we propose to estimate $ENB_{model}$ and $ENB_{all}$ by, respectively, $\text{mean}(\widehat{NB}^*_{model})$ and $\text{mean}(\widehat{NB}^*_{all})$, i.e., average of NBs across the bootstrapped samples. Indeed, these quantities converge to the point estimates of their counterparts in the original sample ($\widehat{NB}_{model}$ and $\widehat{NB}_{all}$) as the number of Monte Carlo iterations increases. The use of bootstrap-averaged values over original sample estimates, however, has some advantages. First, this approach prevents getting occasional negative EVPIs due to Monte Carlo noise. Second, if the process of calculating predicted risks has stochastic components (e.g., imputation of missing predictors), this approach enables the incorporation of uncertainty from such stochastic processes in EVPI calculations. Finally, this approach facilitates the incorporation of prior information via applying rejection sampling to the bootstrap(22).

### *Asymptotic approach based on central limit theorem*

Marsh et al proposed an asymptotic Wald-type inferential method for NB based on deriving the first two moments of the sample distribution of the scaled $\widehat{NB}_{model}$ ($\widehat{NB}_{model}$ divided by outcome prevalence)(23). They showed that the performance of this method, namely in terms of the coverage of the resulting confidence interval, is similar to that of the bootstrap-based method, with the advantage of it being a closed-form estimator and thus not subject to Monte Carlo error. We modify their derivation from scaled NB to unscaled NB, and from specifying a univariate normal distribution for $NB_{model}$ to a bivariate normal (BVN) distribution for $NB_{model}$ and $NB_{all}$:

$$(NB_{model}, NB_{all}) \sim BVN([\widehat{NB}_{model}, \widehat{NB}_{all}], \Sigma),$$

with covariance matrix $\Sigma$ having the following components:

$$var(NB_{model}) = \frac{1}{n}\{P_{True\ Positive}(1 - P_{True\ Positive}) +$$

$$\left(\frac{z}{1-z}\right)^2 P_{False\ Positive}(1-P_{False\ Positive})\ +$$

$$2\left(\frac{z}{1-z}\right)P_{True\ Positive}P_{False\ Positive}\};$$

$$var(NB_{all}) = \left(\frac{1}{n(1-z)}\right)^2 P_0(1-P_0);$$

$$cov(NB_{model},NB_{all}) = \frac{1}{n(1-z)}\left((1-P_0)P_{True\ Positive} + \frac{z}{1-z}P_0 P_{False\ Positive}\right).$$

In estimating the above quantities, $P_0$, $P_{True\ Positive}$, and $P_{False\ Positive}$ are replaced by their sample estimates. With this parameterization, the first term on the right-hand-side of the EVPI equation corresponds to a two-dimensional Unit Normal Loss Integral, for which closed-form solutions are available(24). The computation steps are provided in **Box 2**.

| **Box 2:** Asymptotic method for EVPI computation |
|---|
| 1. Calculate the predicted risk for each individual in the validation sample. |
| 2. Using the predicted risks ($\pi$) and observed outcomes ($Y$), estimate $\widehat{NB}_{model}$, $\widehat{NB}_{all}$, $var(NB_{model})$, $var(NB_{all})$, and $cov(NB_{model},NB_{all})$ for thresholds of interest using the equations in the text. |
| 3. Calculate $A = E\{\max\{0, NB_{model}, NB_{all}\}\}$ using the closed-form equation in Lee et al(24) from the parameters above*. |
| 4. Calculate $EVPI = A - \max\{0, \widehat{NB}_{model}, \widehat{NB}_{all}\}$. |
| *The *mu_max_trunc_bvn* function in the *predtools* R package can be used for this step. |

One limitation of this closed-form solution is that it cannot currently be extended to situations where more than one prediction model is considered. If there are $M$ different models competing with two default strategies, the corresponding $(M + 1) \times (M + 1)$ covariance matrix of incremental NBs can be parameterized using the extension of the equations above. However, the resulting truncated multivariate normal integral might not have a closed-form solution, requiring numerical integration(25).

*Figure 3* demonstrates the EVPI values calculated using the Bayesian bootstrap (red), ordinary bootstrap (blue), and asymptotic (orange) methods across the (0–0.2) threshold (higher thresholds were considered clinically irrelevant and contained <%3 of predicted risks). The asymptotic method for EVPI calculations resulted in an EVPI of 0.0004 at the 0.02 threshold for our case study, while the bootstrap methods both generated an EVPI of 0.0005.

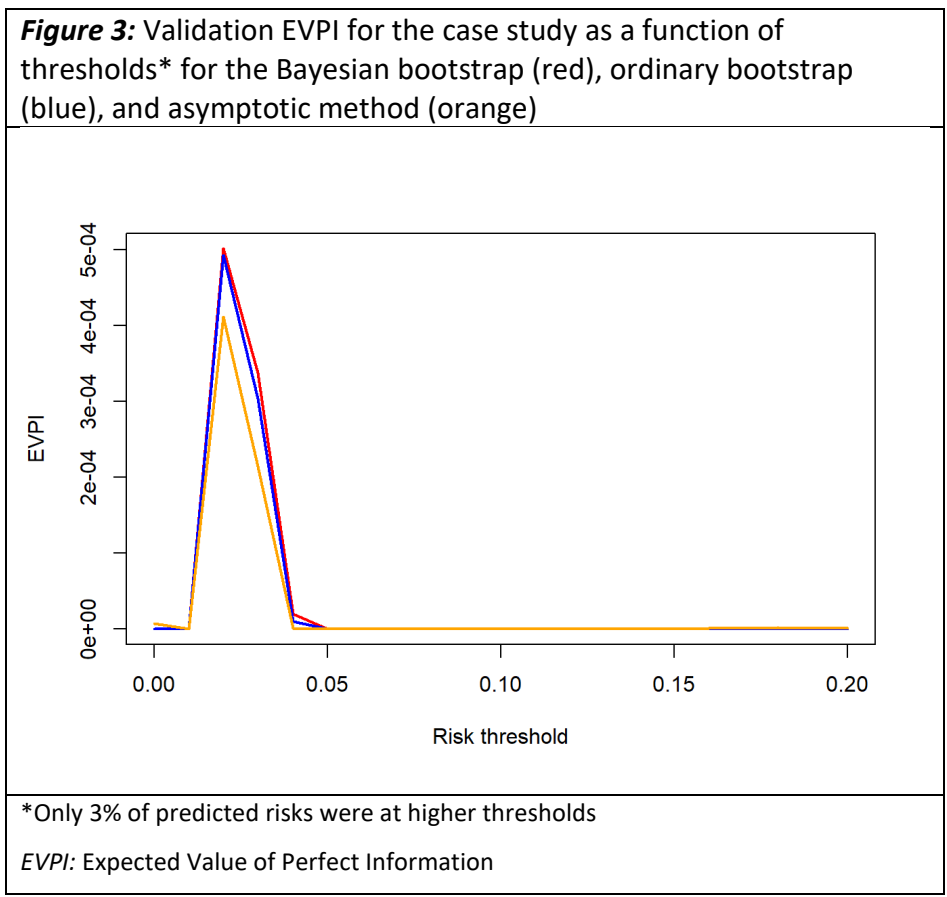

*Figure 3:* Validation EVPI for the case study as a function of thresholds* for the Bayesian bootstrap (red), ordinary bootstrap (blue), and asymptotic method (orange)

*Only 3% of predicted risks were at higher thresholds

*EVPI:* Expected Value of Perfect Information

**Brief simulation studies**

We conducted proof-of-concept simulations to evaluate the consistency of computing EVPI using the proposed algorithms and study how the EVPI changes with sample size.

In the first set of simulations, we considered a simple continuous predictor $X$ with a standard normal distribution, and assumed the outcome-generating mechanism being of the form $logit(P(Y = 1|X)) = -1.55 + 0.77X$. This choice of coefficients creates a scenario where the outcome prevalence is 20% and the correct model for $Y$ has a c-statistic of 0.70 (values that we consider typical in validation studies). We assumed the prediction model happens to be equal to the correct model (returning the correct conditional probability of $Y$ given $X$). We evaluated the EVPI at three exemplary thresholds of 0.1 (low). 0.2 (middle, equal to outcome prevalence), and 0.3 (high). This model has an NB of 0.1176, 0.0575, and 0.0270 at these thresholds. Sample size was varied from 250 to 2000, with doubling in each step.

*Figure 4* provides the results which are the average of 100 simulations. The three computation algorithms generated nearly identical results. As expected, the EVPI declined with larger sample sizes, with an expected pattern of diminishing gains with large samples.

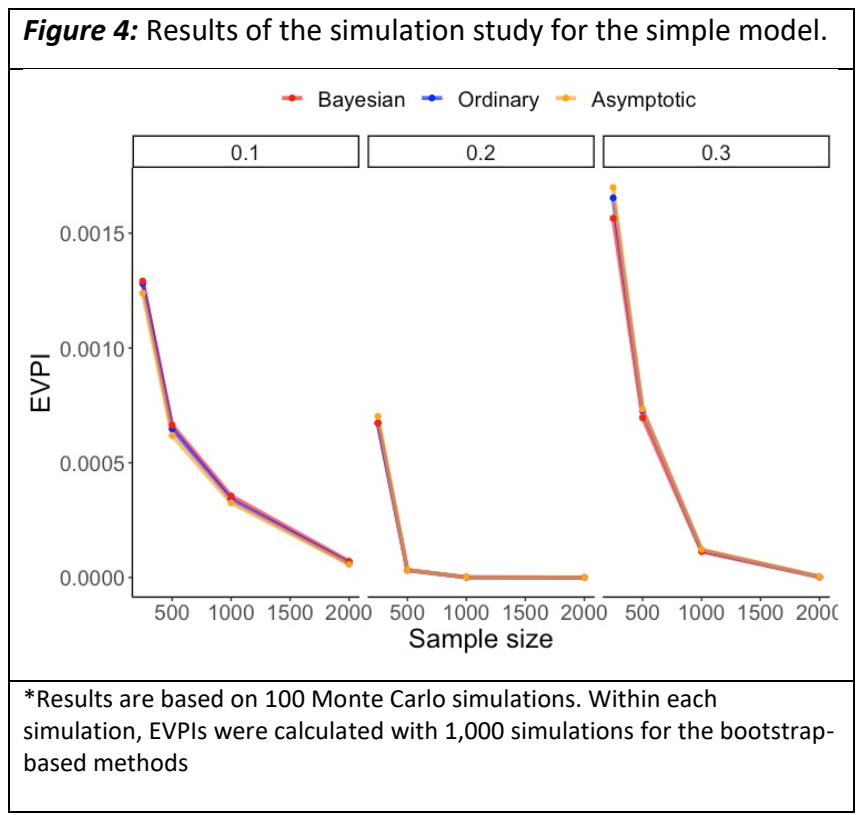

*Figure 4:* Results of the simulation study for the simple model.

*Results are based on 100 Monte Carlo simulations. Within each simulation, EVPIs were calculated with 1,000 simulations for the bootstrap-based methods

*EVPI:* Expected Value of Perfect Information

The second set of simulations was related to the case study. Here, we used an increasingly larger subset of the US sample of GUSTO-I as the external validation dataset. We repeatedly (1,000 times) drew samples without replacement from the entire validation sample, starting from n=250 and doubling it in each step to the maximum size (23,034). We investigated the EVPI at four thresholds of 0.01, 0.02, 0.05, and 0.10. **Figure 5** provides the results on how EVPI changes as a function of sample size for the GUSTO-I study.

*Figure 5:* EVPI values across the range of sample sizes for the three computation methods and threshold values of 0.01 (top left), 0.02 (top right), 0.05 (bottom left), and 0.10 (bottom-right). Red: Bayesian bootstrap; blue: ordinary bootstrap; orange: asymptotic method*

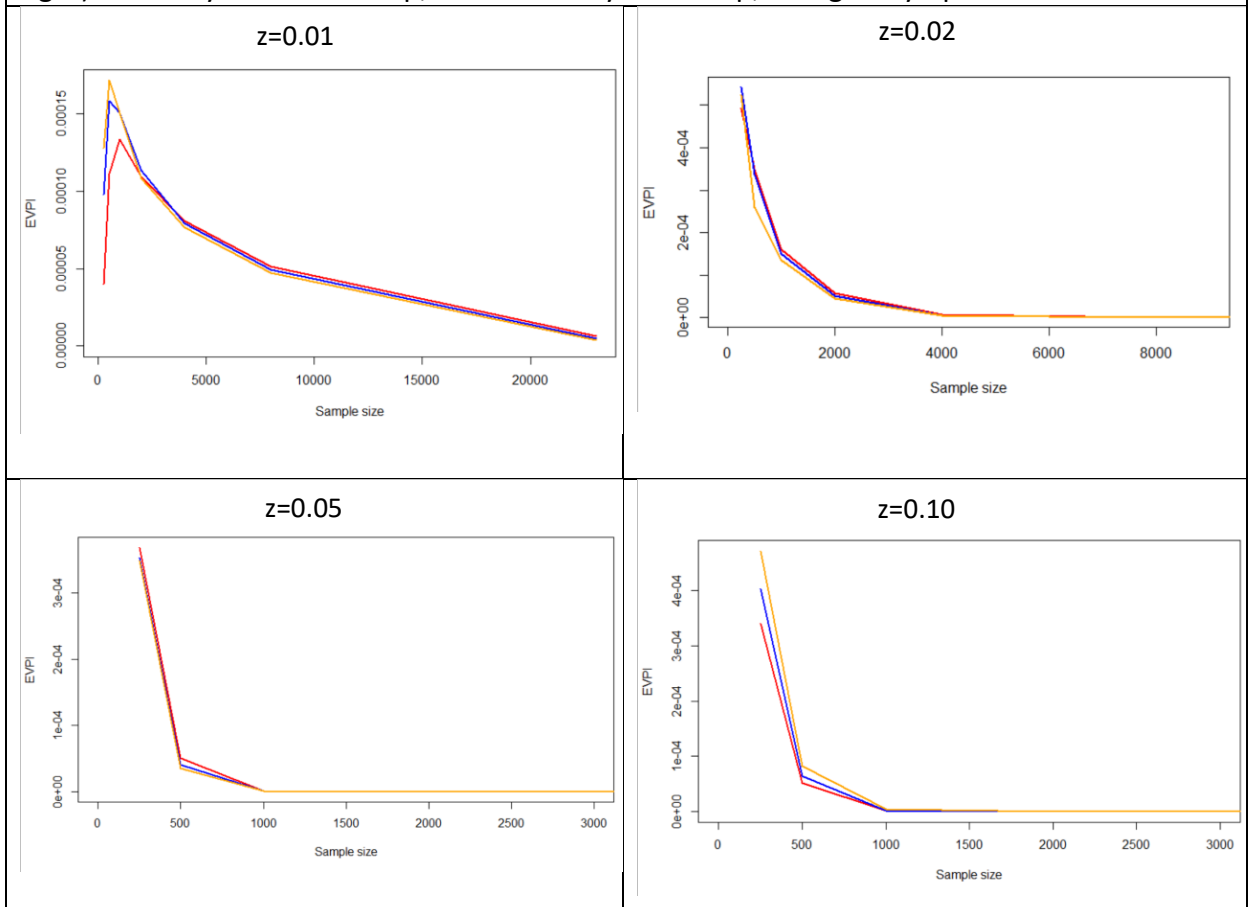

*Results are based on 1,000 Monte Carlo simulations. For the bootstrap methods within each simulation, EVPIs are calculated with 1,000 simulations. The X-axis for three of the panels is truncated as the higher values were all zero.

*EVPI:* Expected Value of Perfect information

Again, all three methods generated similar results. For all thresholds except z=0.01, the EVPI declined with increasing sample size, with the decline being very steep for higher threshold values that were close to outcome prevalence (0.05 and 0.10). With full validation sample the EVPI was 0.000005 for z=0.01, and 0 for other thresholds. The only non-intuitive finding was for z=0.01 at small sample sizes, where the EVPI increased when the sample size increased from 250 to 500 for all three computation methods, and from 500 to 1,000 for the Bayesian bootstrap method. Of note, the 0.01 threshold is significantly lower than outcome prevalence. Only 13% of the original validation sample had a predicted probability below this threshold, and none had an event. This will result in a very low effective sample size to compare the model against default strategies at this threshold.

**DISCUSSION**

We proposed the validation EVPI as the expected loss in NB due to the finiteness of an external validation sample and the associated risk of incorrectly identifying the optimal strategy. We developed algorithms based on bootstrapping and asymptotic methods for EVPI computation. In proof-of-concept simulation studies, we showed EVPI calculation algorithms generated generally consistent results, and EVPI generally behaved as expected, in that with more information (larger sample size) it declined but very large sample sizes were associated with diminishing gain. We interpreted the EVPI in a clinical example for predicting risk of mortality after heart attack. An R implementation of the proposed method is provided in the *predtools* package (https://github.com/resplab/predtools/).

We proposed interpreting the EVPI based on scaling it to the population, as well as comparing it to the NB gain associated with the use of the model. In cost-effectiveness analysis where the payoffs can be converted to net monetary benefit, VoI metrics are typically in monetary units(26). Thus, when scaled to the population, VoI metrics can be compared with the budget of research to objectively inform whether future empirical studies are justifiable(26). However, such calculations involve considering all relevant costs and health consequences of competing strategies over a sufficiently long time-horizon, an approach that most often warrants decision-analytic modeling(27). A main appeal of NB approach in risk prediction is that it uses the information in the validation sample, without having to obtain often jurisdiction-specific evidence or making assumptions on long-term consequences of interventions. While we advocate for such full decision modeling in its due course (e.g., after an impact study has captured the resource use associated with implementing the model at point of care), the EVPI proposed here involves much fewer assumptions and generates results that can be interpreted alongside the results of decision curve analysis. As such, it has the potential to become a standard component of validation studies and provide general guidance on the impact of uncertainty due to the finite validation sample.

Our previously proposed development EVPI captures the expected NB loss due to the distance between the correct (strongly calibrated) model and a candidate model developed using a finite sample(8). Its computation requires modeling $P(Y|X)$ to characterize the distribution of the calibrated risks. Such explicit modeling was not required in the proposed validation EVPI algorithms. This is because the correct model needs not be specified when the goal is to determine if a prespecified model performs better than default strategies. As typical external validation does not involve modeling $P(Y|X)$, this makes validation EVPI calculations easily embedded within the conventional methods used during external validation studies. However, there are instances where VoI analysis during external validation might entail explicit modeling. This might be the case, for example, if one is interested in revising the model in the new population(15). Another instance is where evidence external to the validation sample are available (e.g., information on case-mix from a population-based study). The bootstrap algorithm can be modified to accommodate external evidence, but this approach can become unwieldy if external evidence is multi-dimensional(22). A related context is in multi-center validation studies where hierarchical Bayesian methods can be used to model differences among settings(13,20). Yet another scenario is when the effective sample size is small. This was demonstrated in one of the simulation scenarios with GUSTO-I data, where the low threshold and small sample size resulted in counter-intuitive findings (EVPI declining with a larger sample). In such circumstances, results are overly influenced by a few observations. VoI analysis in such situations might require (parametric or non-parametric) modeling as a form of smoothing or a means to incorporate prior beliefs on the shape of the response function to mitigate the paucity of data. In general, however, the suitability of the available sample in such instances should be carefully considered before embarking on the external validation of a model.

There are several directions for future research in applying VoI to clinical prediction models. Dedicated simulation studies are required to compare more deeply the performance of the EVPI computation methods in various settings. A distinct area of future research is to develop a framework for Expected Value of Sample Information (EVSI) analysis for development and validation of prediction models. While EVPI puts an upper bound on the expected gain with more information, EVSI represents the expected gain in NB associated with an empirical study of a given specification (e.g., sample size), and can thus more specifically guide future research. Further, treatment benefit models that predict an individual's response to a specific treatment are gaining momentum in predictive analytics(28,29), and VoI methods that estimate the expected gain with further development or validation of such models should be developed.

Uncertainty is a fact of life during all stages of predictive analytics including development, validation, implementation, and revision of clinical prediction models. Such uncertainty is conventionally quantified and communicated using classical inferential metrics. VoI combines the probability of incorrect decisions due to uncertainty with the expected loss in clinical utility into a single measure, and therefore provides a complete picture of the consequences of uncertainty(7). VoI metrics, now available for the development and validation phases of risk prediction modeling, have the potential to provide actionable insight on the need for further evidence across the lifecycle of predictive algorithms.